\documentclass[letterpaper,aps,prl,twocolumn,showcaps,showpacs,superscriptaddress,showpacs,amsmath,amssymb]{revtex4}
\usepackage{graphicx,color,longtable}
\usepackage{amssymb,amsmath} 
\def\strutdepth{\dp\strutbox}
\def\nw#1{\strut\vadjust{\kern-\strutdepth\vtop to0pt{\vss\hbox to\hsize
{\hskip\hsize\hskip5pt$\leftarrow$\hss\strut}}}\textcolor{red}{\em #1}}

\begin{document}

\makeatletter
\title{Probing colloidal gels at multiple lengthscales: the role of hydrodynamics}

\author{C. Patrick Royall}
\thanks{paddy.royall@bristol.ac.uk}
\affiliation{HH Wills Physics Laboratory, Tyndall Avenue, Bristol, BS8 1TL, UK,}
\affiliation{School of Chemistry, University of Bristol, Cantock's 
Close, Bristol, BS8 1TS, UK.}
\affiliation{Centre for Nanoscience and Quantum Information, Tyndall 
Avenue, Bristol, BS8 1FD, UK.}
\author{Jens Eggers}
\affiliation{School of Mathematics, University of Bristol, 
University Walk, Bristol, BS8 1TW, UK.}
\author{Akira Furukawa}
\affiliation{Institute of Industrial Science, University of 
Tokyo, 4-6-1 Komaba, Meguro-ku, Tokyo, 153-8505, Japan.}
\author{Hajime Tanaka}
\thanks{tanaka@iis.u-tokyo.ac.jp}
\affiliation{Institute of Industrial Science, University of 
Tokyo, 4-6-1 Komaba, Meguro-ku, Tokyo, 153-8505, Japan.}

\begin{abstract}
Colloidal gels are out-of-equilibrium structures, made up of a rarefied network of colloidal particles. Comparing experiments to numerical simulations, with hydrodynamic interactions switched off, we demonstrate the crucial role of the solvent for gelation. Hydrodynamic interactions suppress the formation of larger local equilibrium structures of closed geometry, and instead lead to the formation of highly anisotropic threads, which promote an open gel network. We confirm these results with simulations which include hydrodynamics. Based on three-point correlations, we propose a scale-resolved quantitative measure for the anisotropy of the gel structure. We find a strong discrepancy for interparticle
 distances just under twice the particle diameter between systems with and without hydrodynamics, quantifying the role of hydrodynamics from a structural point of view. 
\end{abstract}


\maketitle

The formation of a network of arrested material with finite zero-shear viscosity upon slight quenching is among the most striking features of condensed matter \cite{zaccarelli2007,poon2002,coniglio2004,ramos2005sdg}. Gels can be soft or biological materials such as proteins \cite{tanaka2005prl,cardinaux2007}, clays \cite{jabbarifarouji2007}, foods \cite{tanaka2013fara}, hydrogels \cite{helgeson2012} and tissues \cite{drury2003,rose2014}. However a more diverse range of systems including granular matter \cite{ulrich2009}, phase-demixing oxides \cite{bouttes2014} and metallic glassformers \cite{baumer2013} also exhibit gelation. The mechanical properties of gels are influenced by their structure both locally \cite{hsiao2012,sabin2012,zhang2012,zaccone2011} and at a global level through percolation of particles \cite{valadezperez2013} and clusters \cite{kroy2004}, network topology \cite{varrato2012} and confinement \cite{spannuth2012}.

Despite its widespread occurrence, a deep understanding of gelation remains a challenge. In particular, numerical simulations with numbers of particles large enough to reveal the percolating network structure are limited to an approximation in which the effect of the fluid containing the particles is disregarded \cite{holm}. In the colloidal gels we will consider, demixing of the particles into a (colloidal) ``gas'' and ``liquid'' occurs. Spinodal demixing leads to a network of particles 
\cite{verhaegh1997,tanaka1999colloid,manley2005spinodal,lu2008,zaccarelli2008} which undergoes dynamical arrest \cite{testard2011}. The final structure can persist for years  \cite{ruzicka2011}, if the self-generated or gravitational stress is weaker than the yield stress \cite{tanaka2013fara}.

Demixing is driven by effective attractions between the colloidal particles induced by the addition of non-absorbing polymer. Thus although the system is a mixture of three important components --- colloids, polymers and solvent --- in equilibrium we can recast it as an effective one-component system of colloids which experience an attractive interaction whose strength is determined by the polymer concentration \cite{dijkstra2000,likos2001}. However out of equilibrium, hydrodynamic interactions (HI), i.e. interactions between particles mediated by the continuous fluid phase, come into play. These are hard to treat as they are both long-ranged and not limited to pairwise interactions.

\begin{figure}
\includegraphics[width=80mm]{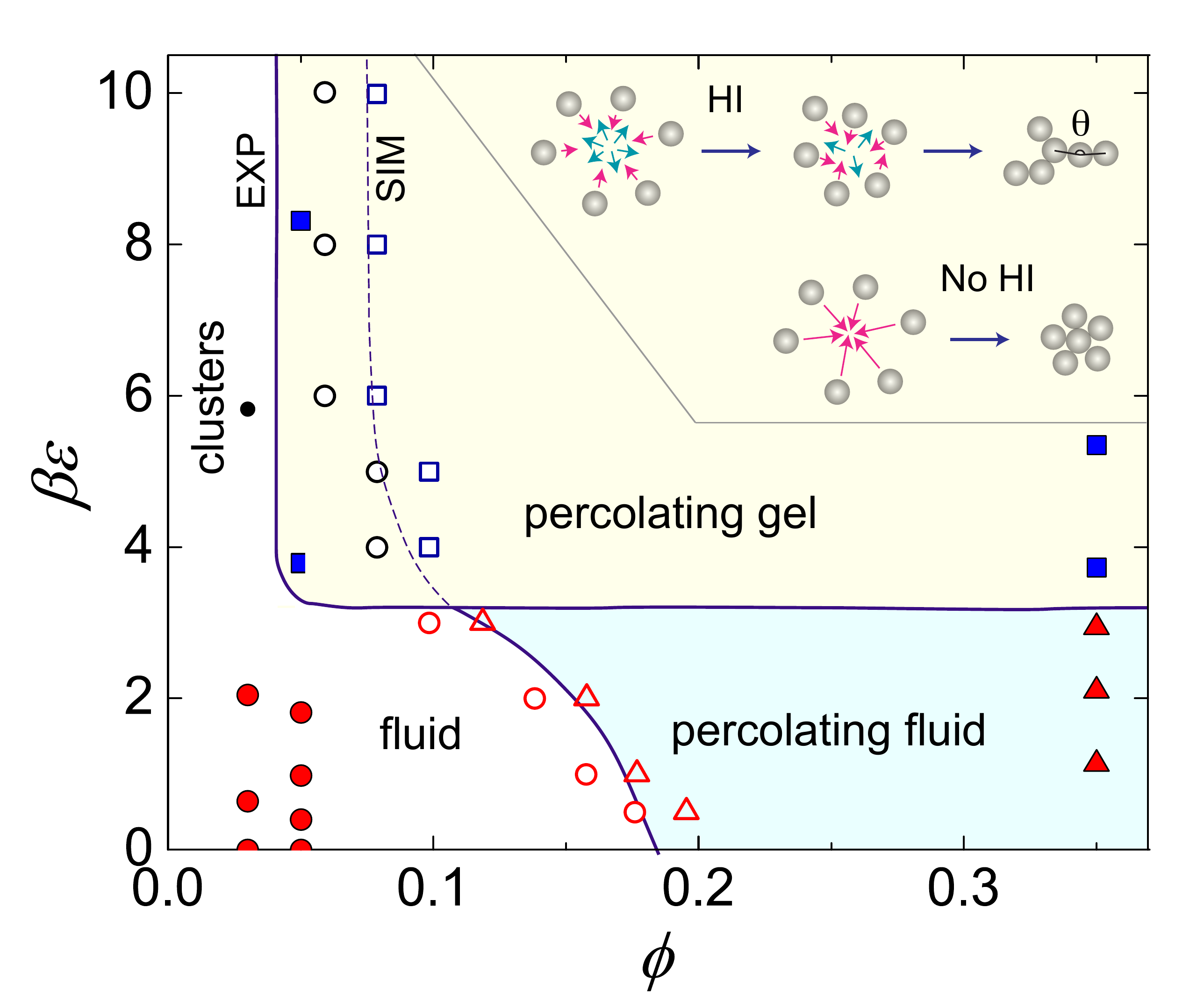}
\caption{(color online) 
State diagram showing gelation in our experimental and simulated system in the $\beta \epsilon$ (potential well depth)-$\phi$ (colloid volume fraction) plane. Phase separation is observed for $\beta \varepsilon \gtrsim 3$. Filled symbols are experimental data, Brownian dynamics simulation data are unfilled symbols. Red symbols are one-phase fluids, circles lie below the percolation threshold, triangles above. Black circles are (non-percolating) cluster fluids. Blue squares are gels.  Inset shows effects of hydrodynamics on colloidal aggregation \cite{furukawa2010}. Top line: hydrodynamic interactions lead to solvent flow (cyan arrows) which influences the motion of the aggregating colloids. The incompressibility of the fluid allows only transverse (rotational) flow, resulting in the formation of an elongated structure rather than a closed one. The degree of elongation is characterized by bond angle $\theta$. Bottom line: no hydrodynamic interactions. Their attractive interactions lead the colloids to aggregate 
(pink arrows) to form a compact structure.} 
\label{figPdMS} 
\end{figure}

In this Letter, we show that the colloid volume fraction required for gelation is reduced by a factor of 1.75 including HI, see Fig. ~\ref{figPdMS}. This occurs although our Brownian dynamics simulations (without HI) are parameterized closely to our experiments \cite{royall2007jcp,royall2008g}. We back up these findings with simulations including HI using fluid particle dynamics  \cite{tanaka2000,tanaka2006viscoelastic} for accessible system sizes. Our results call into question the standard theoretical approach used to describe colloidal systems quantitatively.

We show that HI strongly favor gelation since they inhibit the direct formation of energetically locally favored compact structures, which minimize the potential energy between small groups of $m$ colloids \cite{doye1995,malins2013tcc}. On one hand, the structural mechanism by which the network of colloids forms and becomes rigid has been shown to be condensation into locally favored structures whose geometry underlies the local order of the percolating network \cite{royall2008g}. On the other hand, HI lead to solvent back flow, with the effect of slowing down the condensation of colloids, favoring more open elongated local structures than would be the case without  HI (see Fig. ~\ref{figPdMS} inset) \cite{tanaka2000,tanaka2006viscoelastic,furukawa2010,whitmer2011influence,cao2012hydrodynamic}. This is a consequence of the incompressibility of the liquid solvent, which allows only transverse (rotational) motion of the fluid. Without HI, large closed structures form, favored by minimizing the energy alone.

We used polymethylmethacrylate colloids, sterically stabilized with polyhydroxyl steric acid. The colloids were labeled with 4-chloro-7-nitrobenzo-2-oxa-1,3-diazol (NBD) and had a diameter of $\sigma=2.40$ $\mu$m and polydispersity of 4\% as determined by static light scattering \cite{bosma2002}. The polymer was polystyrene, with a molecular weight of $3.1\times10^{7}$ with $M_w/M_n=1.3$ and solvent was a nearly density- and refractive index-matching mixture of cis-decalin and cyclohexyl bromide. To screen any (weak) electrostatic interactions, we dissolved tetra-butyl ammonium bromide (TBAB) salt to a concentration of 4 mM \cite{royall2007jcp,royall2008g}. We determine the colloid volume fraction $\phi$ and polymer number density $\rho_p$ by weighing along with particle tracking in the case of the former.

We take the polymer volume fraction as $\phi_p=4 \pi R_g^3\rho_p/3$. In these good solvent conditions $R_g\approx160$ nm so the polymer-colloid size ratio is $q \approx 0.18$ \cite{royall2007jcp}. This is much larger than the range of the steric stabilisation layer ($\sim$ 10 nm) previously identified as a criterion for spherically symmetric effective interactions between colloids, which we assume here, neglecting for example any hindrance of rotational diffusion \cite{prasad2003} which might otherwise impact significantly on relaxation dynamics \cite{seto2013}.

We prepare the samples in glass capillaries of internal dimensions 100 $\mu$m $\times$  1 mm $\times$ 1 cm and image the colloidal particles with a Leica SP5 confocal microscope with a resonant scanner and took data at least 20 $\mu$m from the wall of the sample cell. The particle coordinates are tracked in three dimensions (3D) with an accuracy of around $0.03 \sigma$. Samples were prepared 10 minutes prior to imaging and imaged for around 20 minutes. In units of the Brownian time, $\tau_{B}=(\sigma/2)^{2}/6D=3.1$ s, where $D$ is the diffusion constant. Data was therefore taken between $192$ and $578\tau_B$ rafter preparation.  
In our simulations we model the colloid-polymer mixture, using the Morse potential 
\begin{equation}
\beta u(r)=\beta\varepsilon \exp \left[ \rho_0 ( \sigma-r ) \right]   \left(  \exp \left[ \rho_0 (\sigma-r) \right] -2 \right)
\label{eqMorse}
\end{equation}
which we have found to provide an accurate description of colloid-polymer mixtures for size ratios such as we consider here \cite{royall2008g,taffs2010}.

\begin{figure*}[!t]
\includegraphics[width=180mm]{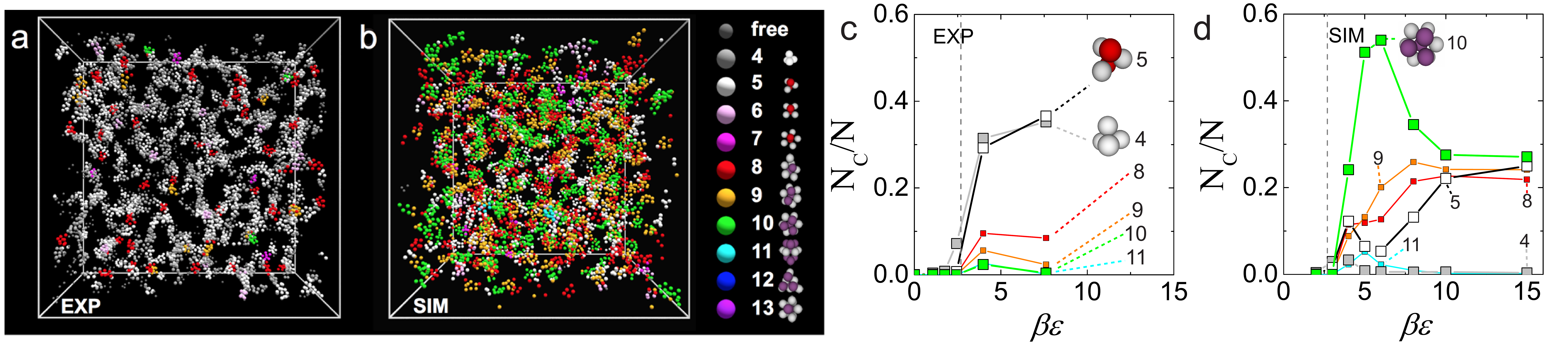}
\caption{(color online) The effect of hydrodynamics on the local structure. (a-b) Snapshots illustrating the effect of hydrodynamics on the local structure of non-equilibrium sticky spheres. Two nearly identical state points are shown, with particles colored according to their local environment as identified with the TCC for $\phi=0.05$, (a) Experiment, $\beta \varepsilon=8.3$. (b) BD simulation, $\beta \varepsilon=8.0$. Cluster topologies are indicated at the right of (b).
(c-d) The effect of hydrodynamics on the local structure for $\phi=0.05$. At low interaction strengths there is little 
clustering. (c) Experimental data are dominated by $m=5$ triangular bipyramids with small quantities of higher-order clusters. (d) BD simulation data show the assembly of many more clusters of more complex geometries. Vertical dashed lines in (c,d) correspond to the (critical) attraction strength required for demixing and line labels refer to the number of particles in the cluster.
\label{figTcc} }
\end{figure*}

\begin{figure}[!t]
\includegraphics[width=70mm]{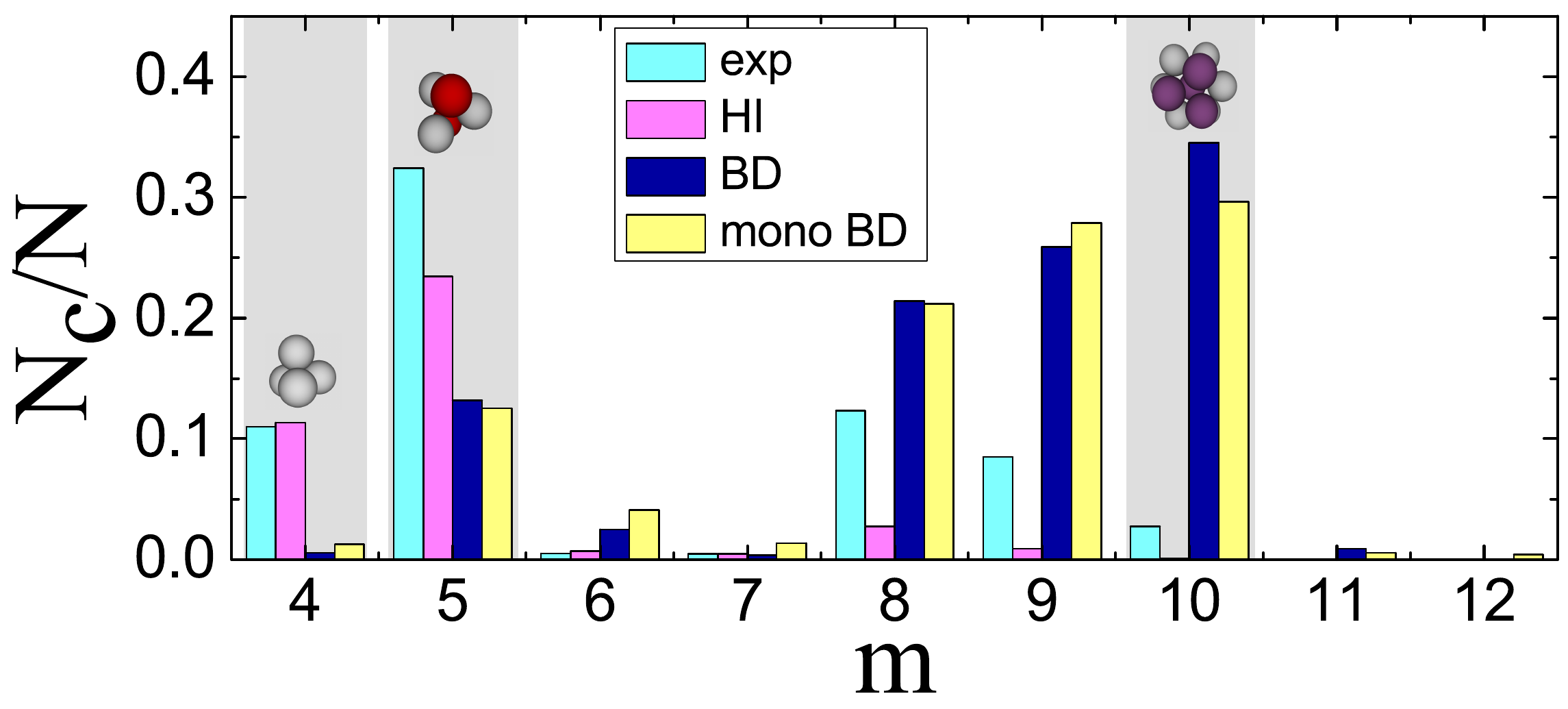}
\caption{(color online) Population of clusters identified with the topological cluster classification. Experimental data corresponds to $\beta \varepsilon=8.3$, for simulations $\beta \varepsilon=8.0$. Simulation data is shown for fluid particle dynamics (HI) for $N=6112$, Brownian dynamics for a larger polydisperse system (BD) and a smaller monodisperse BD system with $N=6112$ (BD mono). Highlighted are $m=4$ tetrahedra and $m=5$ triangular bipyramids favoured in the case of HI and $m=10$ clusters favored by BD simulations. 
\label{figTccCompareHist} }
\end{figure}

The interaction is truncated and shifted at $r=1.4\sigma$ and the range parameter is set to $\rho_{0}=33.0$ \cite{royall2008g}. The effective temperature is controlled by varying the well depth of the potential scaled by the thermal energy $k_{\rm B}T=1/\beta$, $\beta\varepsilon$, which corresponds to the polymer chemical potential in the experiments. This can be estimated from the polymer concentration using Widom particle insertion \cite{widom1963,lekkerkerker1992}. Short-ranged attractive systems can be mapped onto one another \cite{noro2000,foffi2005} via the extended law of corresponding states, which equates the reduced second virial coefficient $B_2^*=2 B_2/(3\pi\sigma^3_\mathrm{eff})$ between two potentials where $\sigma_\mathrm{eff}$ is the effective hard sphere diameter \cite{lu2008,royall2008g,taffs2010}. Here $B_2$ is the second virial coefficient. Knowing the polymer volume fraction in the experiment, we are able to map the effective colloid-colloid interaction to a Morse potential of the same second virial coefficient. We assume that the colloid-colloid interaction has the so-called Asakura-Oosawa form \cite{royall2007jcp}. Further details are provided in Supplementary Material (SM).

Our colloidal particles used in the experiments are slightly polydisperse, which is treated in our Brownian Dynamics (BD) simulations by scaling $r$ in Eq. (\ref{eqMorse}) by a Gaussian distribution in $\sigma$ with 4\% standard deviation (the same value as the experimental system). For details of our BD simulation methods, see \cite{royall2012}. Unless otherwise indicated, runs are equilibrated for $168\tau_B$ (close to the value in the experiments). Data are then sampled for a further $168\tau_B$. The system size was fixed by the box length of $47.13$ $\sigma$ up to $N=40000$ particles (corresponding to $\phi=0.2$). For higher volume fractions, the box was reduced and the system size fixed at $N=40000$. In the simulations we define $\phi=N \sigma^3 \pi / (6 \pi)$. To analyze the local structure, we identify the bond network using the Voronoi construction with a maximum bond length of $1.4\sigma$. Having identified the bond network, we use the Topological Cluster Classification (TCC) \cite{malins2013tcc} to decompose the system into a set of locally favored structures comprised of $m$ particles, which are the minimum energy clusters for the Morse potential \cite{doye1995} and illustrated in Fig. \ref{figTcc}(a) and (b) \cite{malins2013tcc}.

The most striking observation in the state diagram of Fig. \ref{figPdMS} is that experiments form gels (in the sense of a percolating network) even at $\phi=0.04 \pm0.005$ \cite{perc}. By contrast BD simulations form gels only above  $\phi=0.07 \pm0.005$, requiring a much higher volume fraction than in the experiments. We consider different simulation run times and system sizes, neither of which have any significant effect on our results (see SM).

We use the TCC to identify different locally favored structures and consider a state point which typifies the effects of HI on local structure in Fig. \ref{figTccCompareHist}. At $\beta\varepsilon \approx 8$, the experiment exhibits a percolating gel, while BD simulations form isolated (non-percolating) clusters. Under these conditions, the difference in local structure is clear: the experiment (a) is dominated by smaller clusters, mainly tetrahedra and five-membered triangular bipyramids. By contrast the simulation (b) exhibits larger clusters. In Figs. ~\ref{figTcc}(c) (experiment) and (d) (simulation) we consider a range of well depths. We see that over the range relevant to gelation, around 35\%  of the system is in tetrahedra in the experiment, with a further 35\%  is in $m=5$ triangular bipyramids, which are formed by the addition of a particle to the center of a face of a tetrahedron; these two simple clusters dominate the experimental system. This contrasts strongly with the BD simulation data, where much larger, compact clusters based on five-membered rings ($m=8$, 9 and 10) account for over 60\% of the particles in the system. Note that tetrahedra are the smallest rigid unit for spheres in 3D, thus we expect that these represent a limiting case for a dynamically arrested gel structure. Our findings for the local structure are thus consistent with the schematic picture outlined above and also with previous simulation work which showed that HI can suppress the formation of compact clusters \cite{tanaka2000,tanaka2006viscoelastic,furukawa2010,whitmer2011influence,cao2012hydrodynamic}.

To provide further evidence in support of our hypothesis we carry out fluid particle dynamics simulations (FPD). This method includes HI by treating both colloids and solvent as a fluid continuum with a  sufficient viscosity contrast \cite{tanaka2000,tanaka2006viscoelastic}, to realize a situation where viscous dissipation occurs predominantly in the solvent as appropriate to a suspension of solid colloidal particles. Such simulations necessarily describe smaller systems due to the high computational cost. Here we consider a monodisperse system of $N=6112$ particles at a volume fraction of $\phi=0.05$ and attractive well depth $\beta \varepsilon=8$. The results are shown in Fig. ~\ref{figTccCompareHist} and Fig. ~\ref{figTcc}(d). We also carried out a monodisperse BD simulation with $N=6112$, i.e. matched to the FPD parameters. We see that FPD simulations with HI are much closer to the experimental results than either BD result, indicating that the main difference between the BD simulations and experiments is HI, rather than polydispersity or system size.

Thus we argue that the origin of the percolation at low colloid density in the experiment is the lack of compact clusters which lead to a more open structure. These more open clusters arise from hydrodynamic back flow during condensation. While the FPD simulations are very much closer to the experiments than the BD simulations, the agreement is not perfect. This might originate from differences in the system size, as the experimental system is very much larger. Furthermore the boundary conditions imposed by the capillaries in the experiment
 are rather different to the case of of the periodic simulation box. 

\begin{figure}[h]
\includegraphics[width=90mm]{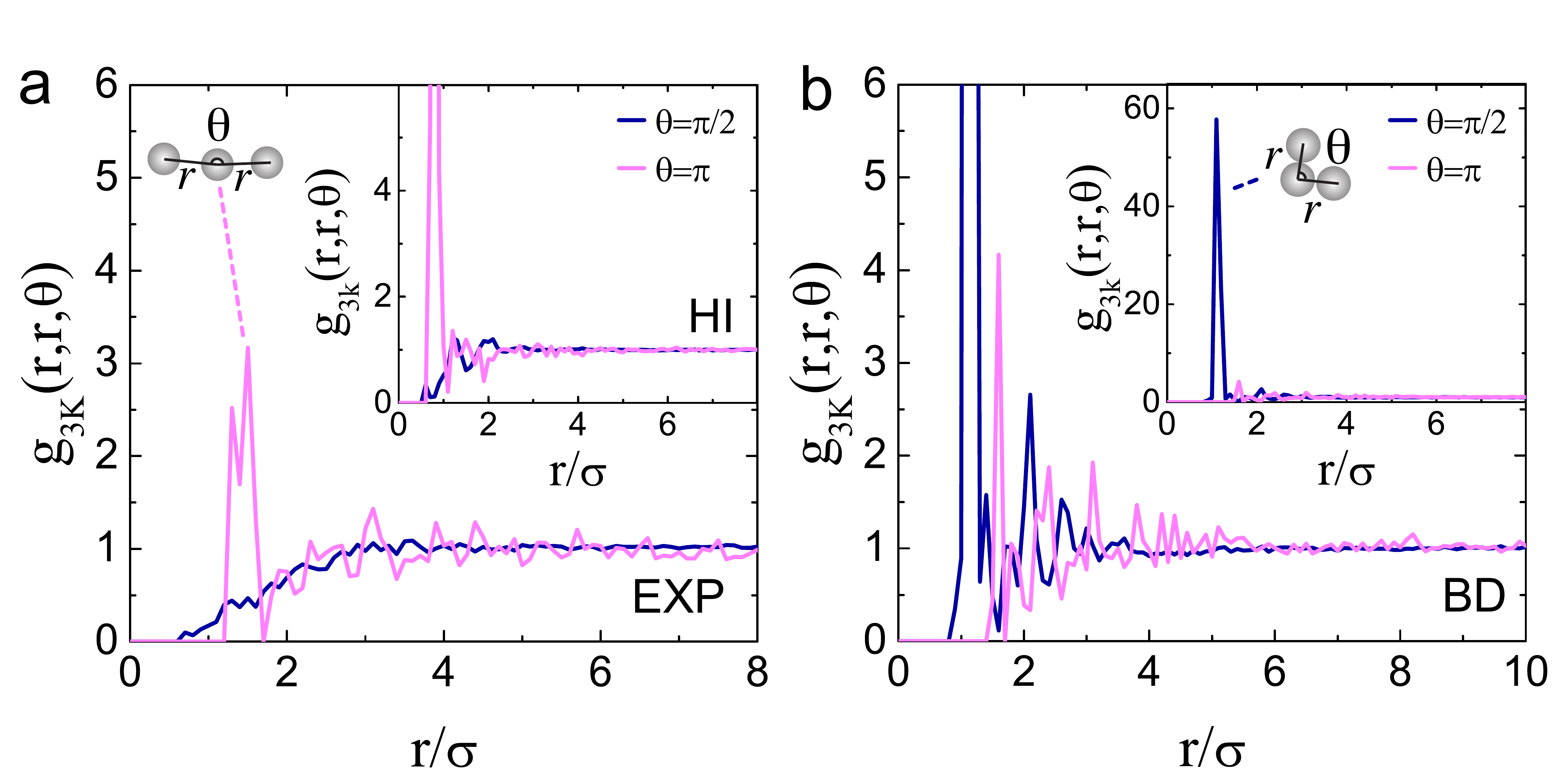}
\caption{(color online) Three-body correlation functions to probe hydrodynamic effects $g_{3k}$ as defined in the text; $\theta$ is the bond angle. (a) Experiment for $\beta \epsilon = 8.3$. Inset: FPD simulation for $\beta \epsilon = 8.0$. (b) BD simulation for $\beta \epsilon = 8.0$. Inset shows the same data expanded to show the full extent of the peak around $\sqrt{2}\sigma$. Simulation data has Gaussian noise (width $0.1\sigma$) added to the coordinates to mimic experimental errors \cite{royall2007jcp}. }
\label{figKirkwood} 
\end{figure}

Having identified the fundamental local structures, we now investigate the lengthscales on which HI lead to thread-like structures, which make up the gel network. Inspired by similar efforts to detect elongated structures in cosmology \cite{gaztanaga2005}, we consider three-body correlations as function of the bond angle. Figure \ref{figPdMS} (inset) indicates that the bond angle distribution is likely to be biased toward bond angles $\theta$ close to $\pi$ compared to the case without HI \cite{tanaka2006viscoelastic,cao2012hydrodynamic}. 
A general three-body correlation function for a triad 
${\bf r},{\bf r}',{\bf r}-{\bf r}'$ may be written as
\cite{hansen}
\begin{equation*}
g_3(\mathbf{r},\mathbf{r}')  = \frac{1}{N \rho^2}\langle \sum_{i \neq j \neq k} 
\delta(\mathbf{r}_k - \mathbf{r}_i + \mathbf{r})\nonumber  \delta(\mathbf{r}_k - \mathbf{r}_j + \mathbf{r}') \rangle\;,
\label{eqG3}
\end{equation*}
where $\rho$ is the number density, and $ijk$ are particle indices. By considering $g_3(r,\theta)$, where $|\mathbf{r}|=|\mathbf{r}'|=r$ and $\theta$ is the angle between  $|\mathbf{r}|$ and $|\mathbf{r}'|$, we probe the angle at different length scales $r$. For simplicity, we compare $\theta=\pi$ and $\pi/2$, a strong signal in the former demonstrating elongated structures typical of a rarefied gel.

We remove the contribution of pair correlations, which can mask the differences in bond angles we seek, using Kirkwood's approximation $g_3({\bf r},{\bf r}')\approx g(|{\bf r}|)g(|{\bf r}'|)g(|{\bf r}-{\bf r}'|)$. We thus plot $g_{3\mathrm{K}}(r,\pi)=g_{3}(r,\pi)/[g(r)g(r)g(2r)]$ and $g_{3\mathrm{K}}(r,\pi/2)=g_{3}(r,\pi/2)/[(g(r)g(r)g(\sqrt 2r)]$ at $\beta \varepsilon \approx 8$ for both experiments and simulations in Figs.~\ref{figKirkwood} (a) and (b) respectively. We see in the simulation data in Fig.~\ref{figKirkwood}(b) that there is a very strong peak at $r \approx \sqrt{2} \sigma$ for $\theta=\pi/2$, corresponding to the compact structure shown in the inset. As expected, no corresponding signal is seen at $\theta=\pi$, which is sensitive to elongated structures.

By contrast, there is a \emph{complete absence} of any peak in the $\theta = \pi/2$ data of the experiment, as seen in Fig.~\ref{figKirkwood}(a). Such a reduction in compact configurations indicates that the presence of HI has a dramatic effect on the geometry of configurations at the three-body level. On the other hand, experiments show a significant peak at around $r\approx 1.7\sigma\lesssim 2\sigma$ for $\theta=\pi$, [cf. Fig.~\ref{figKirkwood}(a)]. This is indicative of chainlike structures with a length scale $\sim 3-4 \sigma$, as expected in the case of HI, which tend to align colloids along the flow, as shown schematically in Fig.~\ref{figPdMS}. There is no detectable signal from clusters such as defective icosahedra, whose length scale is $\sim 3-4 \sigma$ as well, as they do not produce chains sufficiently straight to lead to a peak for $\theta=\pi$. As shown in Fig. ~\ref{figKirkwood}(a) inset, FPD simulations closely follow experiments in their three-body correlations.

In conclusion, we have shown that hydrodynamic interactions have a profound effect on gelation in colloids: the volume fraction above which gelation is found is $\phi \approx 0.07$ for simulations without hydrodynamic interactions whereas $\phi \approx 0.04$ for experiments with HI. Furthermore, we have found that the effects of HI can also be seen from the distinct difference in the particle-level local structures formed upon phase demixing between experiments and simulations without HI, which we have backed up with simulations which include HI. In the absence of HI, compact clusters comprising around ten particles are formed. However those formed in the case of HI at the same volume fraction and attractive strength are predominantly four-membered tetrahedra or five-membered triangular bipyramids. Tetrahedra are the minimum mechanically stable structure. Thus HI promote the formation of tenuous low-density networks. The range over which HI impact local structure is determined with a three-body correlation function. This is sensitive to the spatial extent of anisotropic bond angle corrections. We find that the effects of HI can be observed over distances of $4\sigma$ or around 10 $\mu$m at $\phi=0.05$. We show that, when assessing the low-density limits of gelation, it is crucial to consider hydrodynamics. Thus in designing products based on gels where the concentration of gelling agent is important, one should carry out experiments or simulations which include HI, rather than to rely on Brownian dynamics simulations.

\section*{Acknowledgments}

We gratefully acknowledge stimulating discussions with Tannie Liverpool, Bob Evans and Alex Malins. CPR acknowledges the Royal Society and European Research Council (ERC Consolidator Grant NANOPRS, project number 617266) for financial support and EPSRC grant code EP/H022333/1 for provision of a confocal microscope. Some of this work was carried out using the computational facilities of the Advanced Computing Research Centre, University of Bristol. HT acknowledges Grants-in-Aid for Scientific Research (S) and Specially Promoted Research from the Japan Society for the Promotion of Science (JSPS) for financial support.

\newpage

\section*{Supplementary Material}
\renewcommand{\thefigure}{S\arabic{figure}}

\textit{Mapping interactions between experiment and simulation --- .}
We have previously \cite{royall2007jcp} demonstrated that this system is accurately described by the Asakura-Oosawa interaction between the colloids, which reads

\begin{widetext}
\begin{equation}
\beta u_{\mathrm{AO}}(r)=
\begin{cases}
\infty  & \mathrm{for}  \hspace{12pt} r \leq \sigma \\
- \phi_p^r  \frac{(1+q)^3}{q^3}  \left[ 1-\frac{3r}{2(1+q)\sigma} +\frac{r^{3}}{2(1+q)^{3}\sigma^{3}}  \right] & \mathrm{for} \hspace{12pt} \sigma < r \leq \sigma+\sigma_p \\
0 & \mathrm{for } \hspace{12pt} \sigma+\sigma_p < r \\
\end{cases} 
\label{eqAO}
\end{equation}
\end{widetext}

\noindent Here $\phi_p^r$ is the polymer volume fraction in a reservoir in chemical equilibrium with the experiment. Widom particle insertion \cite{widom1963,lekkerkerker1992} is used to map between reservoir and the experimental system. $\sigma_p=2 R_g$ is the polymer diameter where $R_g$ is the radius of gyration. The Asakura-Oosawa interaction is mapped to the Morse potential as described in the main text.

\begin{figure}
\includegraphics[width=60mm]{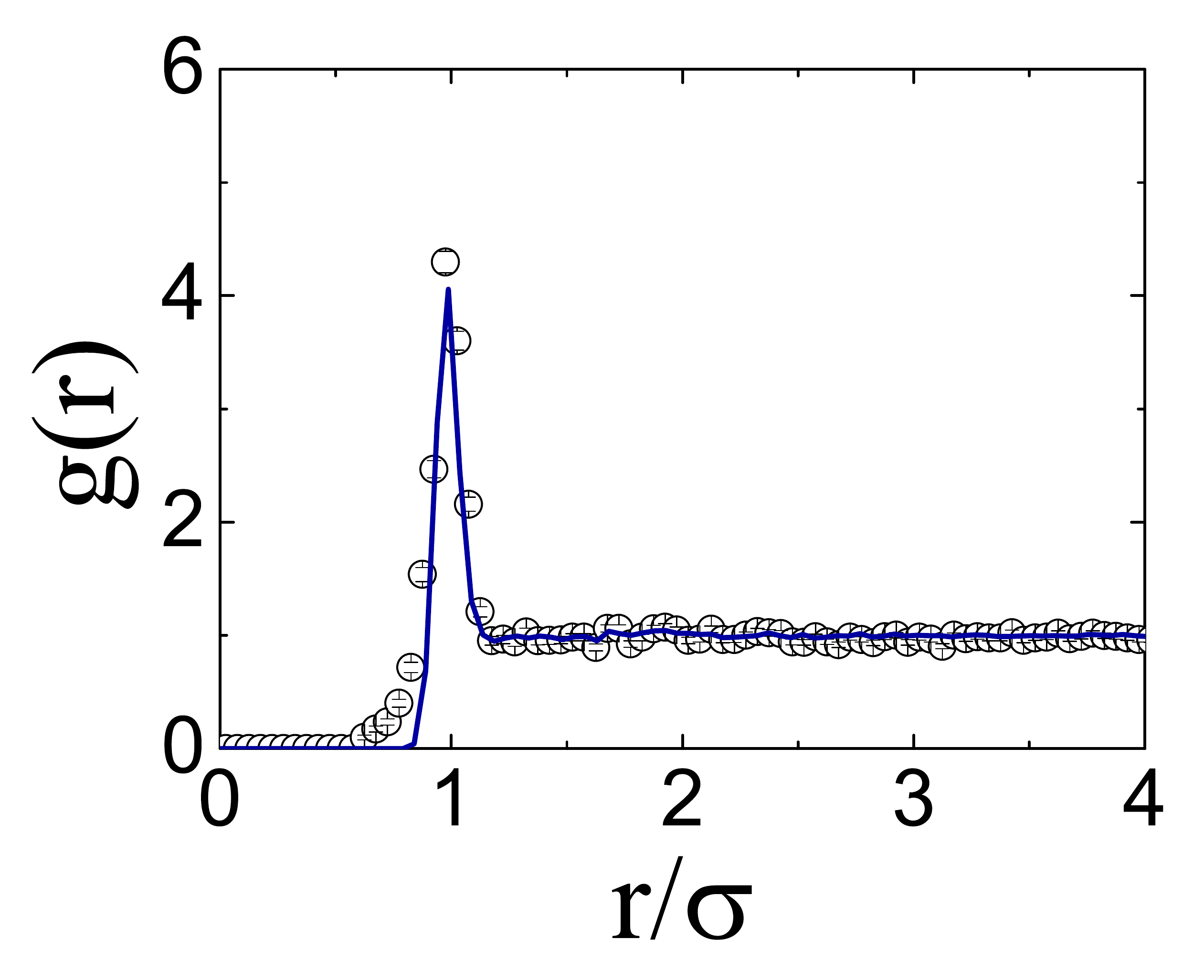}
\caption{(color online) 
Reproducing experiments with simulation. The pair correlation function 
in experiment (circles) is reproduced by Monte Carlo simulations with the 
Morse potential. These have a polydisperse colloid size distribution of the same form as the Brownian Dynamics simulations.
In the simulations the well depth of the Morse potential $\beta \epsilon=1.8$. }
\label{sFigGPhi0117} 
\end{figure}

To investigate the accuracy of our technique in Fig. \ref{sFigGPhi0117} we show the pair correlation function $g(r)$ of a stable fluid reproduced with simulation. A Gaussian distribution 
of standard deviation of $0.03 \sigma$ is added to the coordinates to mimic the particle tracking errors in the experiment \cite{royall2007jcp}. The agreement in the $g(r)$ indicates that the pairwise interactions in the experiment and simulation are very similar.

\begin{figure*}
\includegraphics[width=120mm]{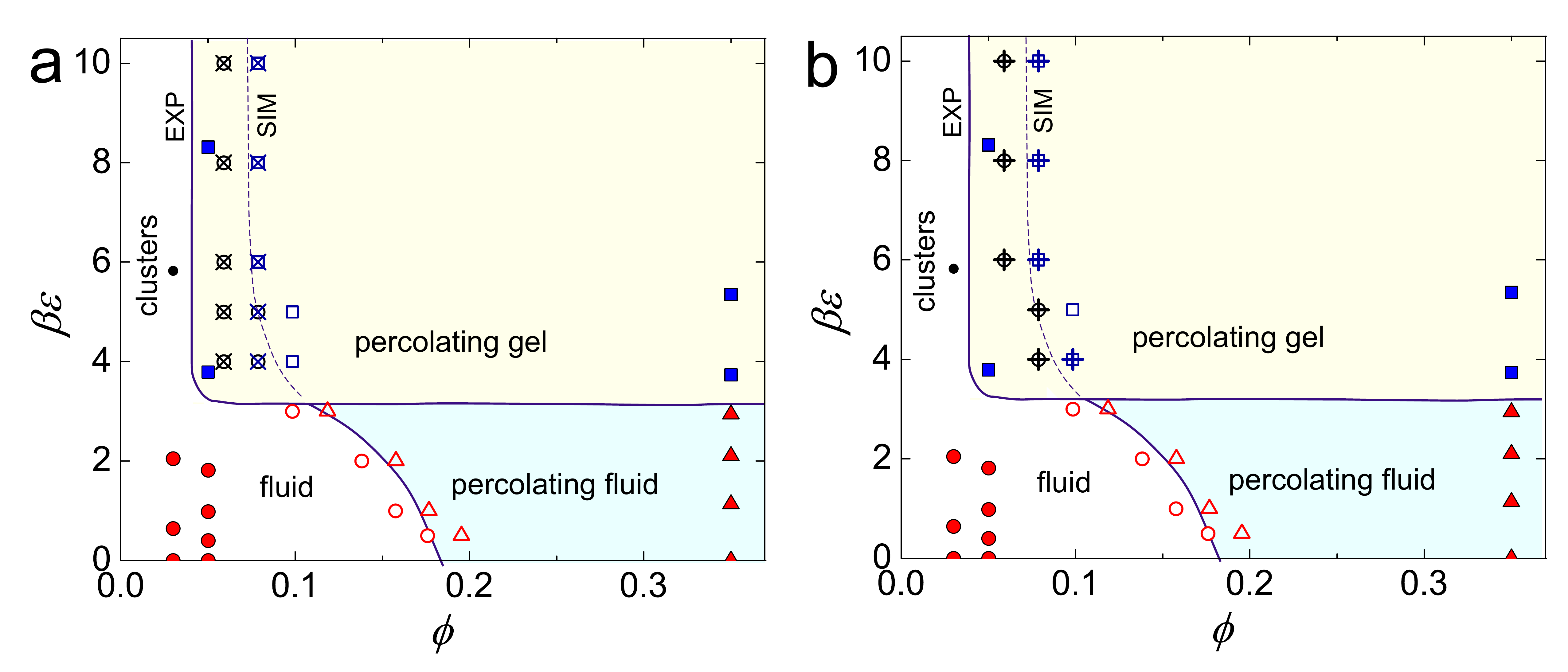}
\caption{(color online) State diagram showing the effects of (a) aging and (b) system size on the measured percolation line in simulations. (a) The effect of ageing is shown by comparing data from simulations which are run for an ``equilibration time'' of $503\tau_B$ rather than $168\tau_B$ as usual. Longer simulations are denoted by ``X''. (b) Larger system sizes. Here we fix $N=50000$ compared to $12000$ for $\phi=0.06$ and $16000$ for $\phi=0.08$. Larger simulations are indicated with crosses. In both panels, all other symbols are the same as Fig. (1) in the main text. }
\label{sFigPdMS} 
\end{figure*}


\begin{thebibliography}{53}
\bibitem[{per()}]{perc}
\bibinfo{note}{In both our experiments and BD simulations, we test for
  percolation by requiring that more than 50\% of configurations have a
  connected cluster of particles which spans the system in at least one
  dimension.}
\end{thebibliography}

\end{document}